# Impact of Rigidity on Molecular Self-Assembly

*Ella M. King[1,2], Matthew A. Gebbie[1,3], Nicholas A. Melosh[1,4]**

*E-mail: nmelosh@stanford.edu

[1]Geballe Laboratory for Advanced Materials, Stanford University, Stanford, CA 94305

[2]Department of Physics, Harvard University, Cambridge, MA 02138

[3]Department of Chemical and Biological Engineering, University of Wisconsin-Madison, Madison, WI 53706

[4]Department of Materials Science and Engineering, Stanford University, Stanford, CA 94305

**Abstract**

Rigid, cage-like molecules, like diamondoids, show unique self-assembly behavior, such as templating 1-D nanomaterial assembly via pathways that are typically blocked for such bulky substituents. We investigate molecular forces between diamondoids to explore why molecules with high structural rigidity exhibit these novel assembly pathways. The rigid nature of diamondoids significantly lowers configurational entropy, and we 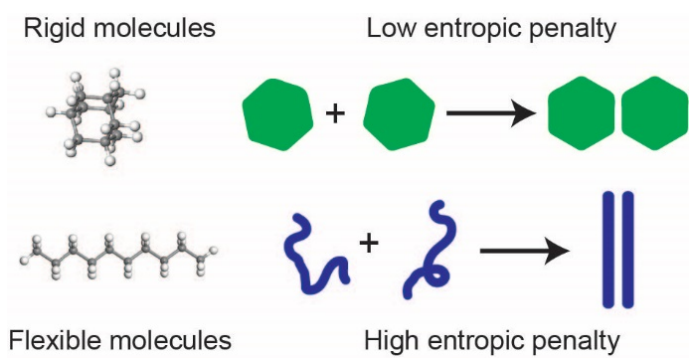 hypothesize that this influences molecular interaction forces. To test this concept, we calculated the distance-dependent impact of entropy on assembly using molecular dynamics simulations. To isolate pairwise entropic and enthalpic contributions to assembly, we considered pairs of molecules in a thermal bath, fixed at set intermolecular separations but otherwise allowed to freely move. By comparing diamondoids to linear alkanes, we draw out the impact of rigidity on the entropy and enthalpy of pairwise interactions. We find that linear alkanes actually exhibit stronger van der Waals interactions than diamondoids at contact, because the bulky structure of diamondoids induces larger net atomic separations. Yet, we also find that diamondoids pay lower entropic penalties when assembling into contact pairs. Thus, the cage-like shape of diamondoids introduces an enthalpic penalty at contact, but the penalty is counterbalanced by entropic effects. Investigating the distance dependence of entropic forces provides a mechanism to explore how rigidity influences molecular assembly. Our results show that low entropic penalties paid by diamondoids can explain the effectiveness of diamondoids in templating nanomaterial assembly. Hence, tuning molecular rigidity can be an effective strategy for controlling the assembly of functional materials, such as biomimetic surfaces and nanoscale materials.

**Introduction**

The self-assembly of hierarchical structures is fundamental to many areas of science and technology.[1–3] Understanding how to predict and tune the assembly of complex materials remains a topic of significant interest,[4] and rationally designing hierarchical assemblies that mimic those present in biological systems remains challenging. Increasingly, researchers are demonstrating strategies to self-assemble complex multifunctional materials that are composed of uniform inorganic building blocks, like nanoparticles,[5–8] to address emerging technological needs.

A key to achieving structural control in nanoparticle assemblies is to control both enthalpic interaction forces via charge and polarity,[6] as well as tuning entropic forces via particle size, shape, and symmetry.[5] Controlling the entropic features of assembly via shape and symmetry has been studied for assemblies of rigid colloids and nanoparticles, where the configurational entropy of the individual particles is often minimal. In contrast, typical small molecules, like linear alkanes, exhibit substantial configurational entropy, which is anticipated to have a large influence on self-assembly processes.

For example, recent work has shown that the most energetically favorable conformation for linear alkanes changes from an all-*trans* linear conformation to a folded back hairpin conformation as the length of aliphatic hydrocarbons increases to above about 17 carbon atoms.[9] Further, computations have shown that the conformational contribution to hydrocarbon formation energies can exceed 1 kcal/mol for linear alkanes with more than 8 carbon atoms,[10] which is comparable to the energy of thermal fluctuations at 295 K. There are even indications that changes in conformational entropy can be a decisive factor that influence the stability and rigidity of cell membranes for ammonia metabolizing extremophiles.[11] Nevertheless, there has not yet been a quantitative study of how large the impact of conformational entropy is on the intermolecular forces that drive molecular assembly.

Here, we explicitly study the influence that molecular rigidity has on self-assembly agents by examining the interplay between molecular flexibility and the distance dependence of intermolecular interactions upon the approach of two molecules. We use diamondoids as highly rigid molecules and linear alkanes as flexible species. Our rationale for comparing diamondoids to linear alkanes as a model system is that these two classes of molecules represent the extremes of high conformational entropy (alkanes) and low conformational entropy (diamondoids), while maintaining the same sp3 chemical bonding, which should maintain similar dispersion interactions between molecules with the same number of carbon atoms.

Experimentally, there are strong indications that understanding diamondoid assembly could reveal novel methods for engineering molecular self-assembly. Recently, diamondoids were used to template the synthesis of metal chalcogenide nanowires with a three-atom cross-section,[12] which occurred due to the presence of a unique assembly mechanism where bulky diamondoid substituents all rotate to the same side of a growing cluster. This in contrast to intuition gained from typical steric interactions, which suggest diamondoids should locate on opposite corners of growing clusters, blocking assembly. The family of diamondoid molecules also serves as a unique test bed for understanding the transition between small molecular systems, such as adamantane ($C_{10}H_{16}$), and nanoparticles, such as pentamantane ($C_{26}H_{32}$).[13] This system of molecules may enable insights gained from nanoparticle assembly to be applied to molecular systems, and discover if the rich entropic behavior in nanoparticles can manifest at the molecular scale.

Thus far, the unique features of diamondoid-based self-assembly were hypothesized to result from strong van der Waals dispersion forces when compared to non-polar molecules of comparable sizes.[14–16] van der Waals dispersion interactions arise primarily from molecular sizes and polarizabilities,[1–3] and both linear alkanes and diamondoids are hydrogen-terminated sp$^3$ bonded molecules.[17–19] Yet diamondoids are more symmetrical than linear alkanes, which are anisotropic molecules, and thus have polarizabilities that depend on orientation. These geometric effects will likely have a larger influence on the magnitude of dispersion interactions in linear alkanes than in diamondoids, which we further explore below.

Notably, linear alkanes and diamondoids also differ significantly in their rigidity (Fig 1). Where diamondoids exhibit only one conformation, linear alkanes exhibit a wide range of conformations.[20,21] We hypothesize that differences in conformational entropy play a critical role in rationalizing the interaction forces that drive diamondoid self-assembly, in line with previous work on entropic forces in nanoparticle self-assembly.[22] Specifically, we compute enthalpic and entropic interactions as a function of intermolecular separation for diamondoids and analogous linear alkanes composed of the same number of carbon atoms.

Molecular Dynamics simulations of two molecules in a thermal bath of Lennard-Jones (LJ) particles[23] were used to measure the pairwise enthalpic and entropic interactions on approach for diamondoids and linear alkanes. For each distance, the center of mass was fixed but molecules were allowed to freely rotate and reconfigure. The LJ particles serve as a low-density thermal bath, minimizing solvent effects to the potential of mean force, such as solvation energy and solvent ordering. The enthalpy was measured from the pairwise interaction energy relative to infinite separation, and the total free energy was calculated via thermodynamic integration. The entropic contribution was then calculated as the difference between the enthalpy and free energy. The comparison between the rigid structures and flexible structures was made by comparing pairs of adamantane and decane molecules, which both have ten carbon atoms, and diamantane pairs to tetradecane pairs, which both have fourteen carbon atoms.

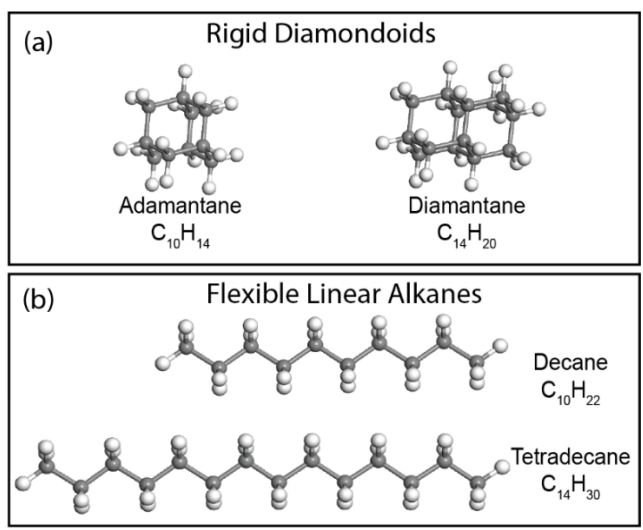

**Figure 1. Molecular structures of diamondoids and alkane counterparts.** (a) The smallest diamondoid molecules, adamantane and diamantane, where the cage-like structures restrict conformational entropy to a single state. (b) Two linear alkane analogues of the diamondoids in (a), decane and tetradecane, where the large flexibility and conformational entropy of linear alkanes strongly influences the free energy of assembly.

This model provides rich insight into diamondoid assembly by enabling the distinction between the distant-dependent contributions of entropy and enthalpy to assembly. We find that differences in configurational entropy and differences in enthalpic dispersion interactions work in opposing ways. As the molecules approach one another, the differences in molecular configurational entropies overcome the differences in enthalpic van der Waals dispersion interactions, providing a molecular explanation for the strong apparent intermolecular interactions between diamondoid molecules during assembly.

**Computational Methods**

Molecular dynamics (MD) simulations were performed using Lammps,[24] which has been shown as a robust method for calculating molecular thermodynamic properties.[25] The simulation cell was a cube of side length 200 angstroms. Periodic boundaries were applied in every direction. The initial condition consisted of a pair of molecules at the center of the cell, with Lennard-Jones vapor evenly dispersed throughout the cell in a regular lattice configuration. The Lennard-Jones molecules were initially spaced 40 angstroms apart from one another in all three spatial directions and were allowed to move freely during the simulation. A vapor was used rather than a liquid solvent in order eliminate solvent-specific effects arising from solvent ordering, while still maintaining ergodicity. For the Lennard-Jones parameters, the energy (epsilon) was set to 0.60, and the distance (sigma) was set to 3.0. This corresponds to a Lennard-Jones potential of the form $U(r) = 4\epsilon(\left(\frac{\sigma}{r}\right)^{12} - \left(\frac{\sigma}{r}\right)^{6})$. These values were chosen based on an example provided by moltemplate, a commonly used molecule-builder for Lammps.[24] The OPLS-aa (all atom) force field was used as it includes both carbon and hydrogen parameters explicitly, which provides greater accuracy, although at the cost of an increased simulation time.

Each simulation consisted of a pair of diamondoid or alkane molecules in a bath of Lennard-Jones particles to ensure that entropy and free energy have well-defined values.[26,27] The centers of mass of the diamondoid or alkane molecules were held at a fixed distance, allowing for infinitesimal movements, using the Lammps "fix momentum" method. The method was only used to fix linear momentum—angular momentum was not preset. Each simulation was run for $10^7$ time-steps at 300 K.

For each pair, at least forty independent simulations were run for a range of intermolecular separations. Data points were gathered at separations spanning from direct molecular contact to a distance of greater than 30 angstroms between each pair of molecules (Figs. 2 and 3). Separation distances of 0.0 Angstrom correspond to the center-of-mass separation of lowest energy contact pairs in solid crystals.[18,28] From the MD simulation, we directly calculated the average pairwise energy, average force between molecules, and center-of-mass separation between the two solute molecules. The force vector was projected onto the vector connecting the centers of mass of the two molecules in order to extract a scalar that preserved the sign of the interaction.

The enthalpy at each distance was calculated from the pairwise interaction energy, as the simulation corresponds to a canonical ensemble with a fixed volume and number of particles.[29] In order to find enthalpy changes, the enthalpy value at a given intermolecular separation was subtracted from the value at infinite separation. The value at infinity was determined by averaging the enthalpy values from the region in which enthalpy ceased to consistently increase or decrease as the distance was changed. The total enthalpy was calculated by summing up interactions between each possible pair of atoms between the diamondoid or alkane molecules. Enthalpic contributions due to interactions with Lennard-Jones particles were excluded from the reported values. Molecules were parameterized using the OPLS-aa force field, which approximates van der Waals dispersion interactions calibrated for nonpolar linear hydrocarbons.

The Helmholtz free energy changes were determined through thermodynamic integration, where the free energy is determined by integrating the ensemble average of the derivative of the potential with respecting to a 'coupling parameter' over a series of values of that coupling parameter.[30,31] The ensemble average is equivalent to an average over time in the long-time limit for ergodic systems such as ours. In our setup, the coupling parameter is distance, so the derivative of the potential is the negative of the force.[27] In order to extract the free energy change, we perform the following integral: $\Delta F = \int \langle -f(r) \rangle dr$ where the angular brackets indicate an average. The integral was performed by applying the Matlab "trapz" method to the force data with respect to the distance data.

Though we explicitly compute Helmholtz free energy changes, these free energy changes are equivalent to Gibbs free energy changes when pressure-volume (PV) work is negligible. Because the system is at constant temperature, volume, and particle number, PV work could only arise from the excluded volume experienced by the solvent particles when the approaching alkane or diamondoid molecules are in closer proximity than the diameter of a solvent sphere.

To demonstrate that this contribution to PV work is negligible, we approximated an upper bound of the excluded volume as a cylinder with a length set to the longest possible axis of a diamantane molecule, 6.90 Å, which corresponds to the cross-diamantane diagonal distance. Notably, diamantane is the largest molecule we study. The radius of the cylinder is the radius of a solvent particle, or 1.68 Å. This gives an excluded volume of 61.38 Å$^3$. We then compute the pressure using the van der Waals equation of state, appropriate for the Lennard-Jones vapor. With no excluded volume, the pressure in the system is 15.6476 kcal/m$^3$. When we consider the excluded volume, the pressure changes to 15.6477 kcal/m$^3$. This difference is significantly smaller than the level of precision in these calculations and is thus neglected. Since the PV work is negligible in our system, we refer to free energy and enthalpy changes within the Gibbs formalism.

Lastly, the entropy change as a function of distance was calculated from the difference between the total free energy change and the enthalpy change for each distance. The entropy given in the plots is the calculated entropy change multiplied by -300K, such that the entropy would be given in the same units as enthalpy and free energy. The sign of the entropy was chosen in order to have negative values indicate attraction.

**Results and Analysis**

Our primary goals are to quantitatively assess if van der Waals dispersion forces in diamondoid assembly differ from those of their linear alkane counterparts, to similarly compare the distance-dependent entropic contribution to assembly, and to gain insight into the molecular origin of the pairwise interactions between rigid molecules. While overall interaction forces are unfavorable for each pair of molecules tested in such a dilute thermal bath (vapor phase), differences in the relative interaction forces can be used to understand how molecular driving forces depend on a balance of entropy and enthalpy. Moreover, using a dilute bath allows us to minimize the influence of solvent-specific effects. Note, however, that these low molecular densities imply that self-assembly would not be favorable, thus we expect the free energy for bringing the molecules together to remain positive. In real systems, solvation energies determine if minimal units actually come together to form larger structures.

*Free Energy of Interaction*

We find that the pairwise free energy change for bringing two molecules from infinite separation to direct contact was consistently more favorable for diamondoids than for linear hydrocarbons of comparable sizes at 300K, as shown in Figure 2. This is in agreement with prior experimental findings demonstrating the effectiveness of using diamondoids to template metal dichalcogenide nanowires[12] and to stabilize unusually long carbon-carbon bonds.[15] Free energies of interaction were repulsive for all pairs of molecules in the dilute simulation conditions, as expected. However, these differences in free energies of the alkane and diamondoid species show how rigidity influences the distance-dependent entropic and enthalpic forces that govern self-assembly.

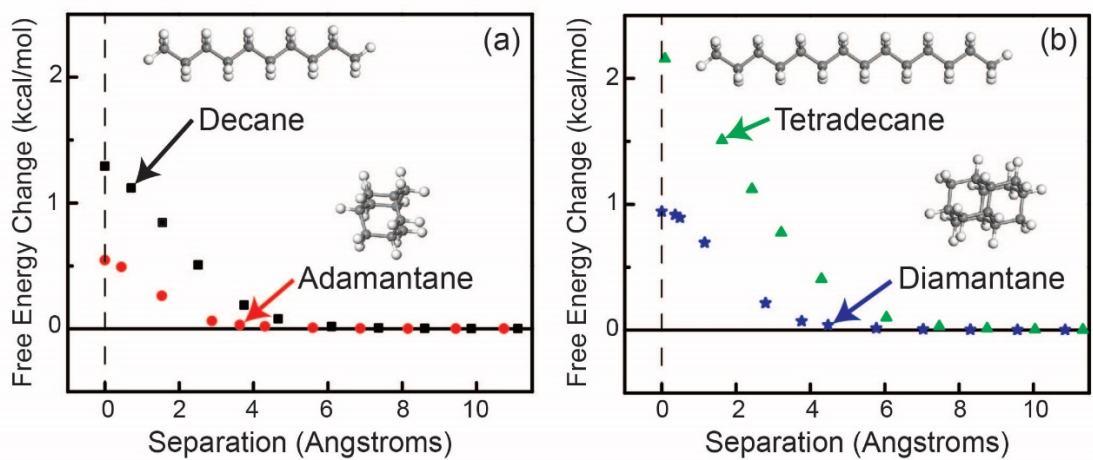

**Figure 2. Free energy change on approach for diamondoids and linear alkanes.** (a) shows the free energy change for a pair of decane molecules (black squares) and a pair of adamantane molecules (red circles), in a LJ thermal bath. (b) shows the free energy change for a pair of tetradecane molecules (green triangles) and a pair of diamantane molecules (blue stars), in a LJ thermal bath. Free energy changes were determined from the forces calculated between the molecules at each separation distance using the thermodynamic integration method explained in the Computational Methods section. Each point represents the free energy change associated with bringing the pair of molecules from infinity to given interatomic separations. The zero-separation value corresponds to molecular contact, and positive slopes in the free energy curve with respect to distance indicate a repulsive force. Hence, it is consistently more favorable to bring diamondoids into contact than linear alkanes of similar sizes under the conditions studied.

The free energy change associated with bringing pairs of molecules into contact from infinite separation was 0.5 kcal/mol for adamantane and 1.3 kcal/mol for decane, a difference of 0.8 kcal/mol between the two pairs. Since the free energy change is positive, this finding suggests that the assembly of adamantane pairs is more favorable than that of decane pairs. For the second diamondoid-alkane comparison, the free energy change associated with bringing tetradecane pairs into contact was 2.2 kcal/mol, as compared to 0.9 kcal/mol for diamantane, a difference of 1.3 kcal/mol. Hence, diamondoid assembly is more favorable than linear alkane assembly under these conditions.

The trends in free energy of assembly follow the same relative trends in the enthalpies of sublimation for each of the molecules tested (Table 1), supporting the applicability of this model for understanding molecular interactions. Pairwise interaction free energies can be used to estimate lattice energies by scaling to account for the number of condensed phase nearest neightbors.[1] The molecules in this study each contain 12 nearest neighbors in the crystalline state.[18,32] The 0.8 kcal/mol difference between the interaction free energies of adamantane and decane molecules corresponds to a simulated difference between the lattice energies of decane and adamantane of about 5 kcal/mol. The measured decane and adamantane sublimation enthalpies differ by 6 kcal/mol, quantitatively agreeing with the simulations to within 20%. The same analysis comparing tetradecane to diamantane yields a predicted lattice energy of 7.8 kcal/mol from the simulated pairwise free energies, which agrees with the measured sublimation enthalpy difference of 8.7 kcal/mol to within 15%.

The computational approach of using MD simulations to determine differences in relative interaction free energies of chemically similar molecules that differ in molecular rigidity thus reproduces trends in experimentally measured thermodynamic data, giving confidence in the simulations. Further, this approach provides reasonable quantitative agreement with experimental data even when comparing between the different molecular species, likely as both have the same $sp^3$-based hybridization.

While we are comparing molecules with the same number of carbon atoms for each of the linear alkane-diamondoid analogues, the molecules have slightly different numbers of hydrogen atoms and electrons. To ensure that the results do not arise primarily from differences in electron number, we normalized the free energies of each pair by number of electrons in each molecule. Doing so collapses the two linear alkane species onto the same line with contact free energy differences of about 0.02 kcal/mol per electron, while the two diamondoids collapse onto a separate line with contact free energy differences of about 0.01 kcal/mol per electron (Fig. S2). This approach maintains the same trends that are discussed and analyzed throughout the main text.

**Table 1. Calculated pairwise changes in free energy, enthalpy, and entropy, and measured bulk phase change enthalpies for decane, adamantane, tetradecane, and diamantane.** Pairwise contact value changes in free energy, enthalpy, and entropy for each molecule were calculated for bringing pairs of molecules from infinite separation to direct contact. Sublimation enthalpies for adamantane and diamantane were measured following the protocol reported in Reference 33. Measurements are discussed in the Supporting Information.[33] All other values were obtained from representative thermodynamic data reported in the National Institute of Standards and Technology (NIST) online Chemistry WebBook.[32] The enthalpy of sublimation for each compound is equal to the sum of the enthalpies of fusion and vaporization: $\Delta_{sub}H = \Delta_{fus}H + \Delta_{vap}H$

| Molecule | Free Energy Change kcal/mol; ($k_BT$ at 300 K) | Enthalpy Change kcal/mol; ($k_BT$ at 300 K) | Entropy Change kcal/mol; ($k_BT$ at 300 K) | Enthalpy of Sub., $\Delta_{sub}H$ (kcal/mol) | Enthalpy of Vap., $\Delta_{vap}H$ (kcal/mol) | Enthalpy of Fus., $\Delta_{fus}H$ (kcal/mol) |
|---|---|---|---|---|---|---|
| Decane | 1.3 (2.2) | -3.2 (-5.4) | 4.5 (7.5) | 19.2 | 12.3 | 6.9 |
| Adamantane | 0.5 (0.8) | -2.4 (-4.0) | 2.9 (4.8) | 13.2 | 12.0 | 2.6 |
| Tetradecane | 2.2 (3.7) | -4.3 (-7.2) | 6.5 (10.9) | 28.1 | 17.1 | 10.8 |
| Diamantane | 0.9 (1.5) | -3.5 (-5.9) | 4.4 (7.4) | 19.4 | - | 2.1 |

*Enthalpic van der Waals Dispersion Interaction*

Enthalpic contributions to the total pairwise interaction free energies are shown in Figure 3. Enthalpy values calculated for linear alkanes were consistently more attractive than those for comparable diamondoids. At contact, decane molecules exhibit an enthalpic attraction of -3.2 kcal/mol, while adamantane molecules exhibit an enthalpic attraction of -2.4 kcal/mol (Fig. 3a). For these 10 carbon molecules, this is a difference of 0.8 kcal/mol. Similarly, we calculate a contact enthalpic attraction of -4.3 kcal/mol for tetradecane and -3.5 kcal/mol for diamantane (Fig. 3c), a difference of 0.8 kcal/mol.

As with the interaction free energies, the trends in calculated pairwise enthalpies agree with the measured thermodynamic data shown in Table 1. When the pairwise enthalpies are scaled to account for the number of nearest neighbors in crystals of each molecule, the calculated enthalpies agree with the measured sublimation enthalpies to within 25% (Table 1). For example, the calculated enthalpy change of decane is 19.2 kcal/mol compared to a measured value of 19.2 kcal/mol, and the calculated enthalpy change of adamantane is 14.4 kcal/mol compared to a measured value of 13.2 kcal/mol.

Further, our calculations are in reasonable agreement with first principles calculations of contact interactions in linear alkanes,[34] given that the final states sampled in our simulation are unlikely to correspond to the single lowest energy pair present in bulk crystals. Specifically, pairwise interaction energies for long axis-aligned contact pairs of hexane dimers range from a value of -1.90 kcal/mol to -4.58 kcal/mol for the most stable dimer.[34] Hence, our calculated value of -3.2 kcal/mol for decane and -4.3 kcal/mol for tetradecane imply that we are sampling contact pairs that are representative of average pairwise contact interaction energies, as opposed to single global minima. Agreement between our calculations and measured thermodynamic properties imply this approach is effective for exploring how conformational rigidity influences molecular assembly.

The observation that linear hydrocarbons consistently exhibit more favorable enthalpic interactions than diamondoids can be explained by steric effects. Since diamondoid molecules are bulkier than linear alkanes, the atoms on two different linear alkane molecules are on average closer together than the atoms on two different diamondoid molecules at a given separation. Intriguingly, recent work has suggested that direct dihydrogen contacts between adjacent C-H bonding groups in cage-like molecules, like diamondoids, are actually stronger than for linear alkanes.[14] That study used first principles calculations that capture subtle differences in chemical bonding, which are not included in our model. Hence, some of the relative enthalpic penalty we calculate may be offset by electronic effects, which could further explain the efficiency of diamondoids in templating self-assembly.[16]

Notably, solely considering the enthalpic contribution to pairwise interactions would suggest that linear alkanes should form more stable contact pairs than diamondoid molecules. However, the total calculated interaction free energies show that diamondoid assembly is favored over the assembly of linear alkanes under these conditions. As discussed below, this observation can be rationalized by explicitly considering the distance-dependent influence of configurational entropy on molecular assembly.

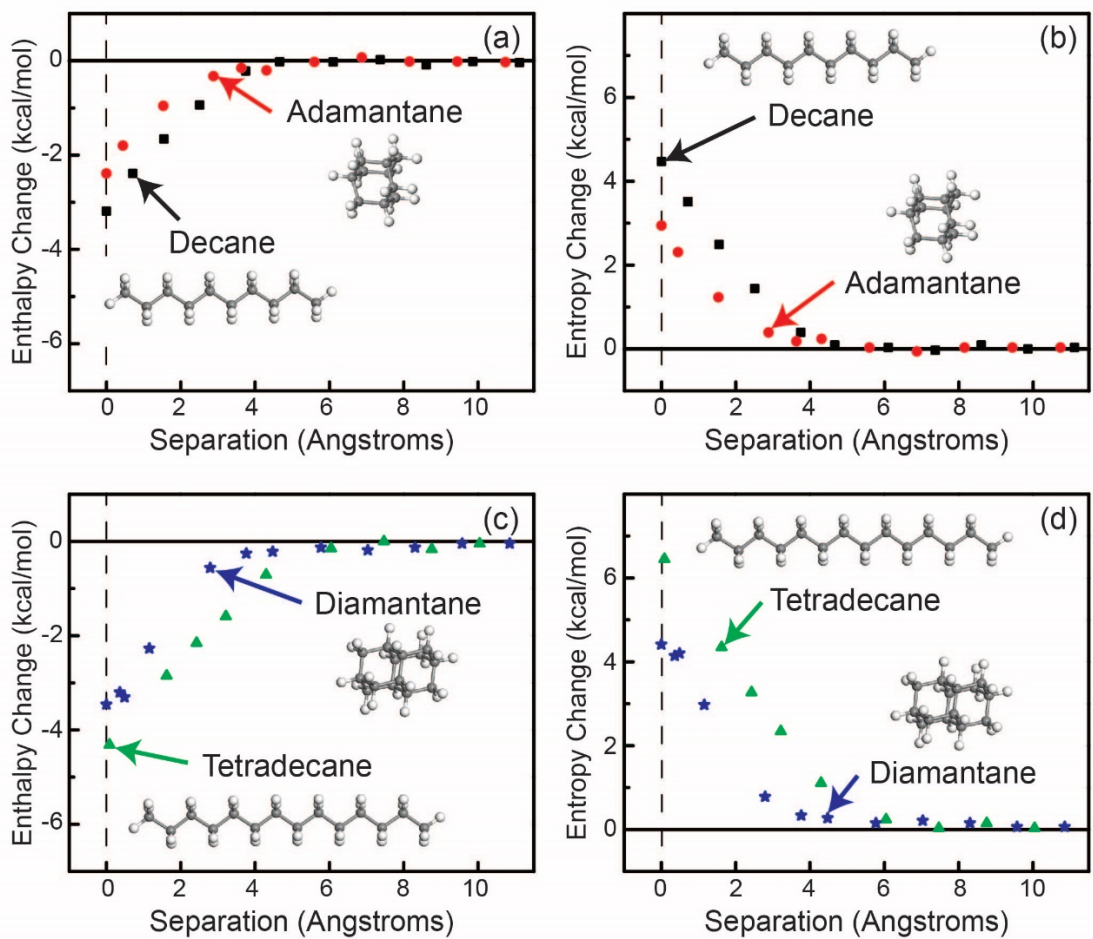

**Figure 3. Enthalpy and entropy change on approach for diamondoids and linear alkanes.** (a) shows the enthalpy change for a pair of decane molecules (black squares) and a pair of adamantane molecules (red circles) and (b) shows $-T\Delta S$, which is the entropy change multiplied by the temperature of 300 K, for a pair of decane molecules (black squares) and a pair of adamantane molecules (red circles) in a LJ thermal bath. Positive values of $-T\Delta S$ indicate an unfavorable decrease in entropy. (c) shows the enthalpy changes for a pair of tetradecane molecules (green triangles) and a pair of diamantane molecules, and (d) shows $-T\Delta S$, which is the entropy change multiplied by the temperature of 300 K, for a pair of tetradecane molecules (green triangles) and a pair of diamantane molecules (blue stars) in a LJ thermal bath. Positive values of $-T\Delta S$ indicate an unfavorable decrease in entropy. Each point represents the change in enthalpy or entropy associated with bringing pairs of molecules from infinity to given interatomic separations. The zero distance corresponds to molecular contact in the crystalline state.

*Entropic Contribution*

      We explicitly examined the influence of configurational entropy on assembly, since we found that diamondoid molecules form contact pairs that are more stable than pairs formed by

linear alkanes, despite the fact that linear alkanes exhibit more favorable enthalpic interactions at contact. Entropy changes were found by subtracting the total free energy change from the enthalpy change. As shown in Figure 3b, the entropic penalty associated with bringing together two decane molecules at 300 K is 4.5 kcal/mol, which is more than 50% higher than the entropic penalty of 2.9 kcal/mol associated with forming adamantane contact pairs. Likewise, the penalty for forming tetradecane contact pairs of 6.5 kcal/mol is nearly 50% higher than the entropic penalty of 4.4 kcal/mol associated with forming diamantane contact pairs (Fig. 3d).

For both the adamantane-decane and tetradecane-diamantane comparisons, differences in entropic penalties exceed differences in enthalpic stabilization. Adamantane enthalpic stabilization at contact is 0.8 kcal/mol less favorable than decane, yet this difference is compensated by the 1.6 kcal/mol reduction in the unfavorable configurational entropy penalty associated with forming contact pairs. For the diamantane-tetradecane comparison, diamantane exhibits a 0.8 kcal/mol lower enthalpic stabilization compared to tetradecane, but this is again offset by a larger 2.1 kcal/mol reduction in the entropic penalty associated with assembly. Hence, while rigid diamondoids have a decreased enthalpic stabilization at contact, this decrease in enthalpic stabilization is offset by an even larger relative entropic benefit during assembly, as the rigidity of the diamondoids minimizes the entropic cost associated with molecular assembly.

Overall, our analysis of the influence of rigidity on pairwise interactions highlights how entropy and enthalpy balance during assembly processes. Intriguingly, the agreement between our simulations and measured phase change energetics implies bulk calorimetry may provide insights into how entropy influences pairwise interactions. Since the vaporization enthalpies are in agreement to within 2% for adamantane and decane (Table 1), sublimation enthalpy differences, which are challenging to measure, primarily arise from differences in enthalpies of fusion. This same observation holds for tetradecane and diamantane, as the enthalpy of vaporization for tetradecane is equivalent to the enthalpy of vaporization for diamantane (calculated from enthalpies of sublimation and fusion). Hence, our findings suggest that the measurement of solid-liquid melting transitions, which are routine calorimetric experiments, may be useful for exploring how functional group rigidity influences pairwise molecular driving forces during self-assembly.

**Discussion**

Bulky nonpolar functional groups, like diamondoids, feature a prominent role in templating unique self-assembly pathways[12] and stabilizing unusual chemical bonding structures, such as one of the longest carbon-carbon bonds observed to date.[15] Such results highlight the fact that London dispersion forces, the attractive portion of van der Waals interactions for nonpolar species,[1] can substantially influence reaction pathways and self-assembly processes. While London dispersion interactions can be viewed as a 'weak' interaction that is often overpowered by repulsive steric interactions, recent studies suggest that bulky nonpolar substituents impart substantial attractive interactions that can govern chemical bonding and molecular assembly.[16]

Rigid nonpolar groups are especially common structure directing motifs, leading prior researchers to explore how chemical bonding in branched and rigid hydrocarbon molecules influences structural stability.[14–16] To date, studies have focused on how bonding influences enthalpic interactions at molecular contact, where it has been noted that induction effects can increase the strength of dihydrogen contacts between adjacent branched and cage-like nonpolar hydrocarbon molecules, relative to linear alkanes.[14] While these results help explain the structure-directing propensity of diamondoids and other cage-like molecules, other researchers have noted that assembly interactions between rigid species like diamondoids and nanodiamonds still appear to be surprisingly strong when compared to linear hydrocarbons, invoking other mechanisms like electrostatic interactions to explain pronounced self-assembly.[35]

By studying how molecular rigidity influences pairwise interactions between nonpolar hydrocarbons, we explore if differences in configurational entropy can also be used to explain the unique assembly properties of diamondoids and other rigid nonpolar species.[16] In examining interaction enthalpies, we find that linear alkanes actually exhibit stronger van der Waals dispersion interactions than diamondoids at all separations: the physical bulk of diamondoids leads to greater net separations between atoms, leading to lowered enthalpic stabilization. Compared to diamondoids, nearly all of the atoms in linear alkanes are closer to the interface between the molecules. Further, linear alkanes have more hydrogen atoms than their diamondoid counterparts.

However, by directly computing the entropic contribution to molecular interactions, we find that the enthalpic penalty diamondoids pay in assembly is offset by a lowered entropic penalty. Near molecular contact, linear alkanes are largely limited to a single conformation. Far from contact, a large number of conformations are available (Fig. 4). By contrast, diamondoids have only one conformation available at all separation distances. As such, alkanes lose a significant amount of conformational entropy when assembling into contact pairs, whereas diamondoids lose no conformational entropy. Moreover, diamondoids are highly symmetric molecules, which reduces rotational entropy. Both factors increase the propensity for diamondoids to from contact pairs, relative to linear alkanes, which helps to explain why diamondoids and other rigid species can be effective as self-assembly and structural stabilization functional groups.[12,15,16]

For example, we find that the assembly of two adamantane molecules is 0.8 kcal/mol more favorable than for two decane molecules, and the assembly of two diamantane molecules is 1.3 kcal/mol more favorable than for two tetradecane molecules. Hence, the conformational entropy penalty paid by these larger linear alkane species during assembly is about 1 kcal/mol. This finding is in agreement with the contributions of conformational entropy to hydrocarbon formation energies, which has been calculated to be about 1 kcal/mol for linear octane.[10] Importantly, the magnitudes of conformational effects exceed the approximately 0.5 kcal/mol energy of thermal fluctuations at 300 K, and thus are large enough to influence molecular assembly.

Indeed, in a recent study of diamondoid-enabled nanowire growth,[12] molecular dynamics simulations were used to show that structures where adamantane substituents rotate to the same side of a growing nanowire core are about 1.2 kcal/mol more stable than configurations with the diamondoid species located at opposite sides of the central growing unit, which would block further nanowire growth. Notably, typical steric considerations imply that large substituents of 10 carbon atoms in size would be located in a 'blocked' configuration to reduce the steric repulsion between bulky species.[3] Our calculations suggest that differences in conformational entropy can be large enough to be the decisive factor explaining the co-location of adamantyl- substituents during these self-assembly processes.

Finally, our comparison of linear alkane moieties to diamondoid analogues with the same number of carbon atoms illustrates that changes in molecular rigidity can impact intermolecular interaction forces with magnitudes exceeding 2 kcal/mol (Table 1). As a reference point, dipole-dipole interaction forces range from about 1 kcal/mol on the weak end up to approximately 5 kcal/mol for strong hydrogen bonds, as are present in water molecules.[3] Hence, molecular interaction forces can be modulated by tuning rigidity with magnitudes approaching the addition of dipolar functionalities or other chemical groups.

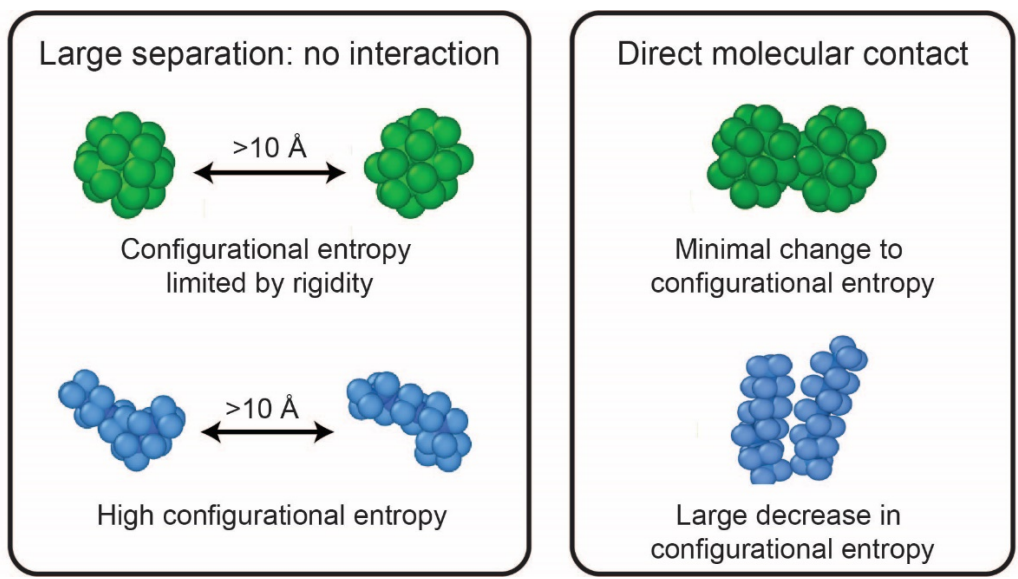

**Figure 4. Molecular rigidity reduces the entropic penalty for assembling into contact pairs.** When the separation distances between pairs of adamantane molecules (green) and pairs of decane molecules (blue) exceed 10 Angstroms, there are no pairwise interactions (left box). At these separations, the conformational entropy of adamantane molecules is already constrained to a single conformation due to high rigidity. Thus, bringing adamantane pairs into contact results in an unchanged conformational entropy (right box). In contrast, decane exhibits significant conformational entropy at large separations as a large number of conformations are accessible via thermal fluctuations. Forming decane contact pairs substantially reduces this conformational entropy, leading to an overall higher entropic penalty associated with self-assembly than when compared to rigid molecules, like adamantane.

**Conclusion**
Our simulations show that the rigidity of diamondoids introduces an enthalpic penalty for assembly, but this penalty is compensated by entropic effects. We conclude that molecular rigidity helps to explain the unique structure-directing assembly behavior that has been observed for diamondoids and should also be applicable to other rigid species. More broadly, our results illustrate that changes in molecular rigidity influence conformational entropy, which can impact assembly processes with magnitudes that approach that of introducing dipoles or other chemical functionalities. Finally, we demonstrate that potential of mean force calculations can be used to calculate the distance-dependent influence of entropy on assembly for small molecules. We anticipate that our approach can be extended to further explore how entropy influences self-assembly behavior, with applications ranging from biomaterials[36] and supramolecular assemblies[37] to nanotechnology[5] and energy materials.[38]

**Acknowledgements**

The authors thank Fei Hua Li, for her invaluable assistance in using the Lammps software, and for her continued advice in the research. We also thank Wei Cai and Yanming Wang for their advice and expertise in the design of the computational approach. This work was supported by the Department of Energy, Office of Basic Energy Sciences, Division of Materials Science and Engineering, under contracts DE-AC02-76SF00515. Portions of this work were performed at the Stanford NanoFacilities (SNF), supported by the National Science Foundation under award ECCS-1542152. M.A.G. acknowledges the funding provided by the Geballe Laboratory for Advanced Materials Postdoctoral Fellowship program at Stanford University. The computational work used resources at the Stanford Research Computing Center.

**Dedication**

This manuscript is dedicated to the memory of Jacob N. Israelachvili, whose creativity, commitment to mentoring, and sense of humor have positively inspired and influenced numerous scientists.

**Supporting Information.** Measurements of Diamondoid Sublimation Enthalpies and Electron Rescaled Free Energy Differences of Diamondoids and Linear Alkanes.

The sublimation enthalpies for adamantane and diamantane were measured following the protocol reported in Reference 33. Briefly, Thermal Gravimetric Analysis (TGA) was used to determine the steady state mass loss of mg-sized samples of each diamondoid under a range of constant temperatures. After determining the steady state mass loss at each temperature, the measured mass loss was used to determine the enthalpy of sublimation via the relationship:

$$\log\left(m_{sub}T^{\frac{1}{2}}\right) = \frac{-0.0522(\Delta H_{sub})}{T} + \left[\frac{0.0522(\Delta H_{sub})}{T_{sub}} - \frac{1}{2}\log\left(\frac{1306}{M_w}\right)\right]$$

Where $m_{sub}$ (μg/min) is the steady state mass loss at temperature T (K), $\Delta H_{sub}$ (J/mol) is the sublimation enthalpy, $T_{sub}$ (K) is the sublimation temperature, and $M_w$ (g/mol) is the molecular weight. Hence, the slope of a linear fit to a plot of $\log(m_{sub}\ T^{1/2})$ versus $1/T$ yields the sublimation enthalpy, $\Delta H_{sub}$.

For both diamondoid species, the protocol involved adding approximately 10 mg of one of the diamondoids to a platinum TGA pan before loading the pan into a TA Instruments Q500 with a platinum reference pan. The total mass was recorded to μg precision using the high sensitivity balance in the TGA instrument. The steady state mass loss rate at different temperatures was then measured via the following protocols:

For adamantane, the temperature was first ramped at a rate of 20 K/min to a temperature of 323 K. The temperature was then held at 323 K for 10 min, and the steady state mass loss rate was determined between 4 and 9 min of holding. The temperature was then ramped up by 10 K increments at a rate of 20 K/min before being held for 10 more min at each temperature. In all cases, mass loss rates were determined from between 4 and 9 min at each constant temperature. Only temperatures where the mass loss was lower than 35% were used to determine the sublimation enthalpy. The $\log(m_{sub}\ T^{1/2})$ versus $1/T$ plot for adamantane is shown in Figure S1a, where the slope yields a sublimation enthalpy of 55.3 kJ/mol (13.2 kcal/mol). This value agrees with the NIST reported value of 59 ± 4 kJ/mol, which is an average of 18 independent values taken from different analytical techniques.

For diamantane, the temperature was ramped at a rate of 20 K/min to a temperature of 343 K. The temperature was then held at 343 K for 10 min, and the steady state mass loss rate was determined between 4 and 9 min of holding. The temperature was then ramped up by 10 K increments at a rate of 20 K/min before being held for 10 more min at each temperature. In all cases, mass loss rates were determined from between 4 and 9 min at each constant temperature. Only temperatures where the mass loss was lower than 35% were used to determine the sublimation enthalpy. The $\log(m_{sub}\ T^{1/2})$ versus $1/T$ plot for adamantane is shown below in Figure S1b, where the slope yields a sublimation enthalpy of 81.2 kJ/mol (19.4 kcal/mol).

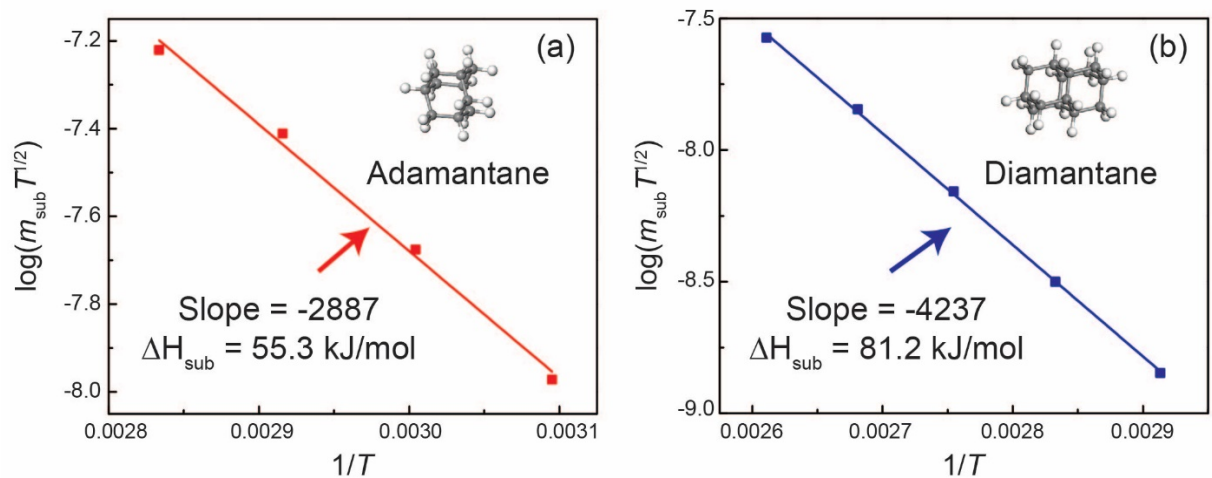

**Figure S1. TGA mass loss data used to determine diamondoid sublimation enthalpies.** Steady state mass loss experiments were performed to determine the sublimation enthalpies of adamantane and diamantane to compare to the MD simulations described in the main text. The experimental protocol was adapted from Reference 32 and is described in the Supporting Methods.

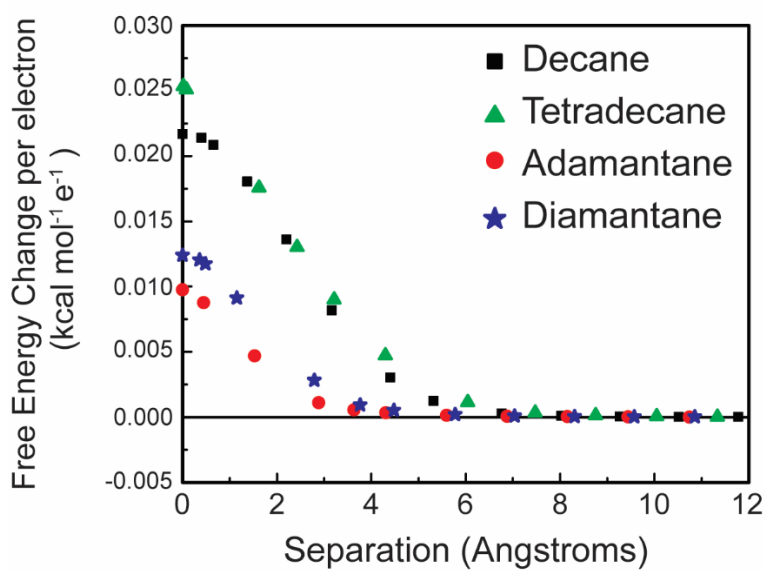

**Figure S2. Electron rescaled free energy change on approach for diamondoids and linear alkanes.** To ensure that the results do not arise primarily from differences in electron number, we normalized the free energies calculated for each pair of molecules by the number of electrons present in one of the molecules. Doing so collapses the two linear alkane species onto the same line with contact free energy differences of about 0.02 kcal/mol per electron, while the two diamondoids collapse onto a separate line with contact free energy differences of about 0.01 kcal/mol per electron. This approach maintains the same trends that are discussed and analyzed throughout the main text.